\documentclass[11pt]{article}

\usepackage{epsfig}
\usepackage{amsfonts,amssymb}

\renewcommand{\theequation}{\arabic{section}.\arabic{equation}}

\newcommand{\s}{\ensuremath{\sigma}}
\renewcommand{\S}{\ensuremath{\Sigma}}

\renewcommand{\th}{\ensuremath{\theta}}

\newcommand{\G}{\ensuremath{{\cal G}}}
\renewcommand{\L}{\ensuremath{{\cal L}}}

\renewcommand{\O}{\ensuremath{{\cal O}}}

\renewcommand{\d}{\ensuremath{{\rm d}}}

\newcommand{\del}{\ensuremath{\partial}}

\newcommand{\half}{\ensuremath{\frac{1}{2}}}

\def\CM{{\mathbb C}}
\def\UM{{\bf 1}}

\def\a{\alpha}
\def\b{\beta}
\def\g{\gamma}

\def\e{\epsilon}

\def\z{\psi}

\def\l{\lambda}
\def\m{\mu}
\def\n{\nu}
\def\o{\omega}

\def\th{\theta}

\def\r{\rho}
\def\s{\sigma}

\def\x{\xi}
\def\z{\zeta}

\def\G{\Gamma}

\def\L{\Lambda}
\def\O{\Omega}

\def\Mul{{\underline M}}
\def\Nul{{\underline N}}

\def\mul{{\underline m}}
\def\nul{{\underline n}}
\def\pul{{\underline p}}

\def\muul{{\underline \mu}}
\def\nuul{{\underline \nu}}
\def\rhul{{\underline \rho}}
\def\siul{{\underline \sigma}}

\newcommand{\be}{\begin{equation}}
\newcommand{\ee}{\end{equation}}

\newcommand{\ba}{\begin{eqnarray}}
\newcommand{\ea}{\end{eqnarray}}

\newcommand{\ns}{\normalsize}

\begin{document}


\begin{titlepage}

\title{
   \hfill{\ns hep-th/0403235\\}
   \vskip 2cm
   {\Large\bf M-theory compactification, fluxes and AdS$_4$}\\[0.5cm]}
   \setcounter{footnote}{0}
\author{
{\ns\large Andr\'e Lukas\footnote{email: A.Lukas@sussex.ac.uk}
  \setcounter{footnote}{3}
  and P.~M.~Saffin\footnote{email: P.M.Saffin@sussex.ac.uk}} \\[0.5cm]
   {\it\ns Department of Physics and Astronomy, University of Sussex}
   \\
   {\ns Falmer, Brighton BN1 9QJ, UK} \\[0.2em] }
\date{}

\maketitle

\begin{abstract}\noindent
We analyze supersymmetric solutions of M-theory based an a
seven-dimensional internal space with $\mbox{SU}(3)$ structure and a
four-dimensional maximally symmetric space. The most general
supersymmetry conditions are derived and we show that a non-vanishing
cosmological constant requires the norms of the two internal spinors
to differ. We find explicit local solutions with singlet flux and the
space being a warped product of a circle, a nearly-K\"ahler
manifold and $\mbox{AdS}_4$. The embedding of solutions into
heterotic M-theory is also discussed.
\end{abstract}

\thispagestyle{empty}

\end{titlepage}


\section{Introduction}

In trying to understand the structure of M-theory and string theory
there has been much effort put into finding supersymmetric solutions
to the classical bosonic equations of motion. As has been known for
some time the requirement of supersymmetry puts a strong restriction
on the form that solutions can take
\cite{gibbons:82,tod:83,candelas:84}. More recently it has
been suggested that G-structures, obtained from spinor bi-linears, may
be used as a way of classifying supersymmetric solutions
\cite{gauntlett:0205}. Since then there has appeared work using these
ideas to find all supersymmetric solutions to various supergravities theories
\cite{gauntlett:0209}--\cite{caldarelli:0307},
as well as work classifying solutions
\cite{gurrieri:0211}--\cite{Behrndt:2004km} and clarifying
the nature of the supercharges \cite{hackett:0306}.

In this paper we aim to revisit the subject of supersymmetric
solutions of M-theory based on an internal space with $\mbox{SU}(3)$
structure and a four-dimensional maximally symmetric
space of Lorentzian signature~\cite{behrndt:0311}, that is, Minkowski space or
${\rm AdS}_4$. The Ansatz we use allows for general fluxes and we use
the machinery of G-structures to analyze the system, constructing
$p$-forms out of spinor bi-linears. However, unlike in
Ref.~\cite{behrndt:0311}, we allow the norms of the two internal
spinors which define the $\mbox{SU}(3)$ structure to be different.
The general supersymmetry conditions are then derived for this setup,
showing that a non-zero cosmological constant requires the norms of 
the two spinor to be different. More specifically, it turns out
that the cosmological constant acts as a source for the particular
spinor bi-linear which governs the relative normalization of the
spinors. We also find that the external flux as well as one of the
three singlets of the internal flux have to be zero in general. Moreover, for
a non-zero cosmological constant we conclude that the vector part of
the internal flux must vanish and that the internal space can be
sliced into six-dimensional half-flat manifolds.

We apply our general results to the case of internal singlet flux
and derive solutions which, in general, are warped products of a circle,
a six-dimensional nearly-K\"ahler manifold and $\mbox{AdS}_4$.
We also find a special case where the nearly-K\"ahler manifold
degenerates to a Calabi-Yau three-fold, and a nearly-K\"ahler example
with vanishing cosmological constant but non-trivial spinor norms.

The plan of the paper is as follows. In Section \ref{sec:convention},
we set up the Killing spinor equation of 11d supergravity and outline
our conventions, Section \ref{sec:ansatz} introduces the general
Ansatz for the solutions considered in this paper, derives the
supersymmetry conditions and analyzes their general properties and in
Section \ref{sec:sf} we study solutions with $\mbox{SU}(3)$ singlet
flux. Finally, in Section~\ref{sec:hw}, we discuss the possible embedding
of solutions into Horava-Witten theory and, as a consistency check, we
verify that the approximate solutions of Ref.~\cite{Witten:1996mz,lukas:9710} satisfy our
equations. Two Appendices summarize technical information: Appendix~\ref{AppA} contains
gamma-matrix and spinor conventions in $D=4,7,11$; and Appendix~\ref{AppB} the formalism
for $\mbox{SU}(3)$ structures in seven dimensions.

\section{Basic equations and conventions}
\label{sec:convention}

We will work with the following bosonic equations of motion of 11-dimensional supergravity
\ba
\label{geom}
R_{MN}&=&\frac{1}{12}\left(F_{MPQR}F_N^{\;\;PQR}-\frac{1}{12}g_{MN}F^2\right),\\
\label{feom}
{\rm d}\star F&=&-\frac{1}{2}F\wedge F\;,
\ea
where $F$ is the four-form field strength. Indices $M,N,\dots = 0,\dots ,9,\natural$ are
curved 11-dimensional indices and we denote their flat
counterparts by an underlined version, that is $\Mul ,\Nul ,\dots = 0,\dots ,9,\natural$.
The above equations of motion should be supplemented by the Bianchi identity
\begin{equation}
 {\rm d}F=0\; .
 \label{Bianchi}
\end{equation}
A solution of the above system of
equations~(\ref{geom})--(\ref{Bianchi}) is supersymmetric if an
11-dimensional Majorana spinor~\footnote{For a summary of our
gamma-matrix and spinor conventions see Appendix~\ref{AppA}.} $\eta$
exists which satisfies the Killing spinor equation \ba
\label{11Dspinor}
\nabla_M\eta+\frac{1}{288}\left[\Gamma_M^{\;\;\;NPQR}-8\delta_M^N\Gamma^{PQR}\right]
  F_{NPQR}\eta&=&0\; ,
\ea
where we use the standard covariant derivative,
\begin{equation}
 \nabla_M\eta=\del_M\eta+\frac{1}{4}{\omega_M}^{\Mul\Nul}\Gamma_{\Mul\Nul}\eta,
\end{equation}
for spinors. Conversely, a solution of the Killing spinor equation~(\ref{11Dspinor})
which, in addition, satisfies the $F$ equation of motion~(\ref{feom}) and the
Bianchi identity~(\ref{Bianchi}) also satisfies the Einstein equation
and is, hence, a supersymmetric solution of the theory. This follows from the integrability
of the Killing spinor equation \cite{gauntlett:0212}.

\section{Dimensional reduction}
\label{sec:ansatz}

In this paper we are interested in supersymmetric solutions of
11-dimensional supergravity which are maximally symmetric in $3+1$
dimensions, and have an ``internal'' seven-dimensional space with
$\mbox{SU}(3)$ structure. This implies that the metric is a warped
product of a four-dimensional maximally symmetric space-time
(Minkowski space or $\mbox{AdS}_4$) with metric $ds_{(4)}^2$ and a
Euclidean seven-dimensional space with $\mbox{SU}(3)$ structure and
metric $ds_{(7)}^2$.  We denote four-dimensional coordinates by ${\bf
x}= (x^\m )$ with curved indices $\m ,\n ,\dots =0,1,2,3$ and their
flat counterparts $\muul ,\nuul ,\dots =0,1,2,3$.  Seven-dimensional
coordinates ${\bf y}=(y^m)$ are labeled by curved indices $m,n,\dots
=4,\dots ,9,\natural$ and their flat cousins are underlined, as usual.  The
11-dimensional metric can then be written as
\ba
\label{11metric}
{\rm d}s^2_{(11)}&=&e^{2A({\bf y})}{\rm d}s^2_{(4)}+e^{2B({\bf y})}{\rm d}s^2_{(7)}\; ,
\ea 
where $A=A({\bf y})$ is the the warp factor and $B=B({\bf y})$ is a gauge
degree of freedom which will be fixed later by a
convenient choice. The most general Ansatz for the four-form $F$ compatible
with our basic assumptions is given by
\ba
F_{\mu\nu\sigma\rho}&=&f({\bf y})\epsilon_{\mu\nu\sigma\rho},\label{Fint}\\
\hat{F}_{mnpq}&=&\textnormal{arbitrary},\label{Fext}
\ea
where $f=f({\bf y})$ is an arbitrary function of the internal coordinates and $\hat{F}$
denotes the internal part of the form $F$.

\vspace{0.4cm}

In order to manipulate the Killing spinor equation we need to
decompose the 11-dimensional gamma-matrices. Using the four- and
seven-dimensional gamma-matrices $\{\g_\muul\}$ and $\{\g_\mul\}$
introduced in Appendix~\ref{AppA} we define their ``curved'' counterparts
$\{\g_\m\}$ and $\{\g_m\}$ with respect to the vielbeins associated to
$\d s_{(4)}^2$ and $\d s_{(7)}^2$, respectively. The 11-dimensional
gamma-matrices can then be written as
\begin{equation}
 \Gamma_\mu=e^A(\gamma_\mu\otimes 1),\quad\Gamma_m=e^B(\gamma\otimes \gamma_m),
\end{equation}
where $\g$ has been defined in Eq.~(\ref{g5}). 

\vspace{0.4cm}

Finally, we need to find the most general Ansatz for the spinor $\eta$ in our
case. The main building blocks for this Ansatz are the two seven-dimensional
Majorana spinors $\e_{(1)}$ and $\e_{(2)}$ which exist on the internal space
owing to the $\mbox{SU}(3)$ structure~\footnote{For a review of the $\mbox{SU}(3)$
structure formalism for seven-dimensional manifolds see Appendix~\ref{AppB}.}.
The most general Ansatz for $\eta$ based on these two spinors is given by 
\begin{equation}
 \eta=\theta_{(1)}\otimes\epsilon_{(1)}+\theta_{(2)}\otimes\epsilon_{(2)},
\end{equation}
where $\theta_{(1)}$ and $\theta_{(2)}$ are two four-dimensional Majorana
spinors. These two four-dimensional spinors can be related by 
\ba
\theta_{(2)}=(C+D_\mu\gamma^\mu+\half E_{\mu\nu}\gamma^{\mu\nu}+iF_\mu\gamma\gamma^\mu+iG\gamma )
         \theta_{(1)},
\ea
where $C$, $D_\m$, $E_{\m\n}$, $F_\m$ and $G$ are functions of ${\bf y}$. Due to maximal
symmetry $D_\m$, $E_{\m\n}$ and $F_\m$ must vanish and we are left with
\ba
 \theta_{(2)}=(C+iG\gamma_5)\theta_{(1)}\; .
\ea
After a suitable redefinition of $\e_{(1)}$ and $\e_{(2)}$ the spinor Ansatz
can then be written as
\ba
\label{spinorAnsatz}
\eta&=&\theta\otimes\epsilon_{(1)} + i\gamma\theta\otimes\epsilon_{(2)},
\ea
with a four-dimensional Majorana spinor $\theta$.
This has the same form as the spinor Ansatz in
Refs.~\cite{kaste:0303,behrndt:0311,dall:0311} except for one
crucial difference - we do not set the norms of $\epsilon_{(1)}$  and $\e_{(2)}$ equal,
that is, in the language of Appendix~\ref{AppB}, we allow $\l$ to be different from
one. It is, of course, possible to work with spinors $\e_{(1)}$ and $\e_{(2)}$ of
the same norm but then we would have to introduce an arbitrary function in the
Ansatz~(\ref{spinorAnsatz}) in order not to loose any generality. In the
following, we will work with the spinor Ansatz~(\ref{spinorAnsatz}) and unconstrained
spinors $\e_{(1)}$ and $\e_{(2)}$. As we will see this has important consequences
when we analyze the conditions for unbroken supersymmetry.

To complete the spinor Ansatz we need to specify the properties of the four-dimensional
Majorana spinor $\theta$. In the case of Minkowski space $\theta$ is, of course,
simply a constant spinor. In the case of $\mbox{AdS}_4$, we start with the Killing
spinor equation
\begin{equation}
 \nabla_\mu\theta=-\frac{i}{2}\Lambda_1\gamma_\mu\gamma\theta +\half\Lambda_2\gamma_\mu\theta\; ,
\label{4Dspinor}
\end{equation}
where $\L_1$ and $\L_2$ are cosmological constants and $\nabla_\m$ denotes the
covariant derivative with respect to $\mbox{d}s_{(4)}^2$. They imply that the Ricci tensor
associated to $\d s_{(4)}^2$ is given by
\begin{equation}
 R_{\mu\nu}=-3(\Lambda_1^2+\Lambda^2_2)g_{\mu\nu}\; .
\end{equation}
It is known~\cite{lu:9805,behrndt:0311} that spinors $\theta$ which satisfy Eq.~(\ref{4Dspinor})
exist for the two cases $\L_1\neq 0$, $\L_2 =0$ and $\L_1=0$, $\L_2\neq 0$.
We will cover both cases (as well as possible more general ones) by keeping
$\L_1$ and $\L_2$ arbitrary and working with the full Eq.~(\ref{4Dspinor}). 
Also note that the cosmological constants $\L_i$ can be functions of the
internal coordinates ${\bf y}$, a possibility we shall allow for as well.

\vspace{0.4cm}

We are now in a position to evaluate the Killing spinor equation~(\ref{11Dspinor})
for our Ansatz. For the ``internal'' and ``external'' part we find, respectively,
\ba
\label{internalSpinor}
\nabla_m\zeta_\pm&=&-\half\gamma_m^{\;\;n}\del_nB\zeta_\pm+\frac{i}{12}fe^{B-4A}\gamma_m
                   \zeta_\pm\\\nonumber
                  &~&\mp\frac{1}{288}e^{-3B}F_{npqr}\gamma_m^{\;\;npqr}\zeta_\pm
                    \pm\frac{1}{36}e^{-3B}F_{mnpq}\gamma^{npq}\zeta_\pm,\\
\label{externalSpinor}
\frac{1}{288}e^{A-4B}F_{mnpq}\gamma^{mnpq}\zeta_\pm&=&
   \mp\frac{i}{6}fe^{-3A}\zeta_\pm \mp\half e^{A-B}\del^mA\gamma_m\zeta_\pm
   \pm\frac{i}{2}\Lambda_1\zeta_\mp-\half\Lambda_2\zeta_\mp\; ,
\ea 
where the complex spinors $\z_\pm$ are defined by
\begin{equation}
 \z_\pm = \e_{(1)}\pm i\e_{(2)}\; .
\end{equation}
Eqs.~(\ref{internalSpinor}) and (\ref{externalSpinor}) must be satisfied if a
solution is to be supersymmetric and we will analyze these equations in detail below. 

\subsection{Internal relations.}

We begin by re-writing the internal equation~(\ref{internalSpinor}) in terms of the
forms~(\ref{biXi})--(\ref{biXi3Tilde}) constructed as spinor bi-linears. A
straightforward but somewhat tedious calculation leads to
\ba
\label{diff1}
\d(e^{-A}\Xi)&=&0,\\
\label{diff2}
e^{-2A}\d(e^{2A}\tilde{\Xi})&=&-2i\Lambda_1 e^{-A+B}\Xi_{(1)}-2\Lambda_2 e^{-A+B}\Xi_{(1)},\\
\label{diff3}
\d(e^{A+B}\Xi_{(1)})&=&0,\\
\label{diff4}
e^{-3A-2B}\d(e^{3A+2B}\Xi_{(2)})&=&-e^{A-3B}*F+3\Lambda_1 e^{-A+B}\tilde{\Xi}_{(3)+}
                                          -3\Lambda_2 e^{-A+B}\tilde{\Xi}_{(3)-},\\
\label{diff5}
e^{-2A-3B}\d(e^{2A+3B}\tilde{\Xi}_{(3)-})&=&2\Lambda_1*\Xi_{(3)},\\
\label{diff6}
e^{-2A-3B}\d(e^{2A+3B}\tilde{\Xi}_{(3)+})&=&-e^{-3B}\tilde{\Xi}F+2\Lambda_2 e^{-A+B}*\Xi_{(3)}.
\ea
where the subscripts on $\Xi$ refer to the rank of the form. 
The calculation is made less laborious with software designed to perform gamma matrix algebra
\cite{gran:0105}.
We can draw some immediate
conclusions from these relations. Given that $\Xi_{(1)}$ and $\tilde{\Xi}$ are both real,
Eq.~(\ref{diff2}) implies that
\begin{equation}
 \L_1=0\; .
\end{equation}
Hence, only four-dimensional Killing spinors of the second type as parameterized by
$\L_2$ are (potentially) consistent with supersymmetry. In the following we will drop
all $\L_1$ terms in the above relations, and set $\L_1=0$ in the four-dimensional
Killing spinor equation~(\ref{4Dspinor}). 

From Eq.~(\ref{diff1}) we conclude that
\ba
\Xi&=&e^A\; ,\label{norm}
\ea
up to an irrelevant constant of proportionality which we have set to unity for
simplicity. It will be useful to parameterize the relative normalization $\l$
of $\e_{(1)}$ and $\e_{(2)}$, as defined in Eq.~(\ref{norm12}), in terms of
an angle $\chi$ by setting
\ba
\lambda&=&\frac{\cos(\chi)}{1+\sin(\chi)}\; .\label{lchi}
\ea
Note that, from Eq.~(\ref{norm3}), we have $|\l | =|\e_{(1)}|/|\e_{(2)}|$.
Equal norms of the spinors $\e_{(1)}$ and $\e_{(2)}$, therefore, correspond
to the values $\l =\pm 1$, or $\chi =n\pi$, for any interger $n$.
From Eq.~(\ref{norm}) and the definitions (\ref{biXib}), (\ref{biXibtilde}),
(\ref{zeta}) of $\Xi$, $\tilde{\Xi}$ and $\z_\pm$ one finds the norms
$|\e_{(1)}|$ and $|\e_{(2)}|$ then satisfy
\ba
\Xi&=&|\epsilon_{(1)}|^2+|\epsilon_{(2)}|^2=e^A,\label{Xi}\\
\tilde{\Xi}&=&|\epsilon_{(1)}|^2-|\epsilon_{(2)}|^2=e^A\sin(\chi)\label{Xit}\; .
\ea
These relations will allow us to eliminate the implicit dependence
of the supersymmetry equations on the spinor norms in favour of
the warp factor $A$ and $\l$ (or $\chi$). We should stress here that the parameters
$\L_i$ act as a source for the scalar $\tilde{\Xi}$ in Eq.~(\ref{diff2}). By setting
the norms of the spinors $\epsilon_{(i)}$ to coincide, as in Ref.~\cite{behrndt:0311},
the scalar $\tilde{\Xi}$ and, hence, the cosmological constant is forced to vanish.
However, as we have mentioned ealier, it is unnecessary to set these spinor norms equal
and the cosmological constant can only be kept non-zero if one allows $|\e_{(1)}|\neq |\e_{(2)}|$.

\subsection{External relations.}

We now turn to the external supersymmetry equations~(\ref{externalSpinor}).
Using Eq.~(\ref{structure1}), the spinor $\e_{(2)}$ can be eliminated and,
with the help of relations (\ref{structure2})--(\ref{structure5}),
the number of gamma matrices in each term can then be reduced to zero or one.  
Since $(\e_{(1)},\g_m\e_{(1)})$ are linearly independent in spinor space 
one can then obtain a set of bosonic equations
\ba
&&e^{A-4B}\hat{F}\lrcorner\left( J\wedge J
+2\Psi_+\wedge V\right)
=12\left(\L_2-\lambda e^{A-B}V\lrcorner dA\right),\\
&&\lambda e^{A-4B} \hat{F}\lrcorner (\Psi_-\wedge V)
=-2fe^{-3A},\\
&&e^{A-4B}\left( J\wedge V - \Psi_-\right)\lrcorner \hat{F}
=-2f\lambda e^{-3A}V+6\lambda e^{A-B}dA\lrcorner J,\\
&&\lambda e^{A-4B}\left(2\Psi_+\lrcorner \hat{F}
-\hat{F}\lrcorner(J\wedge J)V -(V\lrcorner F)\lrcorner (J\wedge J)\right)
=12\left( e^{A-B}dA+\lambda\Lambda_2 V\right).
\ea
which is equivalent to the external supersymmetry conditions~(\ref{externalSpinor}).
Here, $V$ is a one-form, $J$ a two-form and $\Psi$ is a complex three-form.
These forms are related to the spinor bi-linears $\Xi$ by~(\ref{biXib})--(\ref{biXib3tilde})
and, together, they characterize the $\mbox{SU}(3)$ structure of the
seven-dimensional internal space - as reviewed in Appendix~\ref{AppB}.

At this point it is convenient to decompose the four-form $\hat{F}$
into $\mbox{SU}(3)$ representations, following Ref.~\cite{dall:0311}.
It turns out that $\hat{F}$ contains three $\mbox{SU}(3)$ singlets,
denoted by $Q$, $c_1$ and $c_2$; two vectors $Y$ and $W$ in $({\bf
3}+{\bf\bar 3})_{{\rm SU}(3)}$; a two-form ${\cal A}$ in ${\bf
8}_{{\rm SU}(3)}$; and a three-form $U$ in $({\bf 6}+{\bf\bar
6})_{{\rm SU}(3)}$. This decomposition is explained in Appendix~\ref{AppB}
and the various forms and their properties are summarized in Table~\ref{tab:1}.
The explicit $\mbox{SU}(3)$ decomposition of $\hat{F}$ is given in
Eq.~(\ref{formAnsatz}). With this decomposition of the four-form we find the
external relations become
\ba
\label{ext1}
 e^{A-4B}\left( 4c_1+Q\right)&=&6\left(\lambda e^{A-B}dA\lrcorner V-\Lambda_2\right),\\
\label{ext2}
2\lambda c_2e^{A-4B}&=&fe^{-3A},\\
\label{ext3}
e^{A-4B}\left( W-Y+2c_2V\right)
     &=&-f\lambda e^{-3A}V+3\lambda e^{A-B}dA\lrcorner J,\\
\label{ext4}
\lambda e^{A-4B}\left( (4c_1-Q)V+2W\lrcorner J\right)
          &=&-6\left(e^{A-B}dA+\lambda\Lambda_2 V\right).
\ea
We now make use of the internal co-ordinate freedom and set $B=0$, for
convenience. The inner product of (\ref{ext3}) and (\ref{ext4}) with $V$ (using
some of the relations~(\ref{su3r1})--(\ref{JwV}))
can be compared with (\ref{ext2}) and (\ref{ext1}) which immediately
leads to
\begin{equation}
 f=0\; ,\qquad c_2 =0\; . 
\end{equation}
Hence, we conclude that the external part of the flux~(\ref{Fext}) as
well as the singlet part $c_2$ of the internal flux vanish.
Further, by taking the inner product of (\ref{ext3}) with $J$ and comparing with (\ref{ext4})
we find the two vector parts $Y$ and $W$ of the internal flux are proportional. More
precisely, we have
\ba
\label{yweqn}
Y&=&(1-\lambda^2)W\; .\label{YW}
\ea

\subsection{Summary of supersymmetry conditions}
\label{sec:summary}

It is useful to pause for a moment and summarize the supersymmetry conditions
subject to the simplifications which we have found so far. We have seen from
the internal equations that only the cosmological constant $\L_2$ may be allowed,
and that the normalization of the spinors $\e_{(i)}$ can be expressed,
via Eqs.~(\ref{Xi}), (\ref{Xit}), in terms of the warp factor $A$ and
their relative normalization $\l$. The latter can be conveniently
parameterized by an angle $\chi$ as in Eq.~(\ref{lchi}). From the external
equations we have found that $f=c_2=0$ and that $Y$ and $W$ are proportional,
as specified by Eq.~(\ref{YW}). Further, we adopt the gauge choice $B=0$ from
here on.

The external supersymmetry conditions then simplify to
\begin{eqnarray}
 c_1 &=& -\frac{1}{4}\sin (\chi )Q-\frac{3}{2}e^{-A}\L_2\label{e1}\\
 dA &=& \frac{1}{6}\cos (\chi )QV-\frac{\l}{3}\s\label{e2}\; ,
\end{eqnarray}
where $\s$ is defined by
\begin{equation}
 \s = W\lrcorner J\; .\label{sigma}
\end{equation}
The internal relations can be expressed in terms of the
$\mbox{SU}(3)$ structure $(V,J,\Psi )$ as well, using the
relations~(\ref{biXib})--(\ref{biXib3tilde}) as well as
the spinor normalizations fixed by Eqs.~(\ref{Xi}), (\ref{Xit}).
We find
\ba
\label{diff2b}
e^{-2A}\d\left[ e^{3A}\sin(\chi)\right]&=&-2\Lambda_2 \cos(\chi)V,\\
\label{diff3b}
\d\left[ e^{2A}\cos(\chi)V\right]&=&0,\\
\label{diff4b}
e^{-3A}\d\left[ e^{4A}\cos(\chi)J\right]&=&-e^A\star\hat{F}-3\Lambda_2
     \left[\Psi_--\sin(\chi)J\wedge V\right],\\
\label{diff5b}
\d\left[ e^{3A}(\Psi_--\sin(\chi)J\wedge V)\right]&=&0,\\
\label{diff6b}
e^{-2A}\d\left[ e^{3A}\cos(\chi)\Psi_+\right]&=&-e^A\sin(\chi)\hat{F}-
      \Lambda_2 \left[2\sin(\chi)\Psi_+\wedge V+J\wedge J\right]\; ,
\ea
where here, and in the following, the Hodge star refers to the seven-dimensional
internal space with metric $ds_{(7)}^2$. The internal flux~(\ref{Fint}) now takes
the form
\begin{equation}
 \hat{F}=-\frac{1}{6}QJ\wedge J+J\wedge{\cal A}+\left[V\wedge J +(1-\l^2)\Psi_- \right]\wedge W
   -c_1\Psi_+\wedge V+V\wedge U\; .
 \label{Fspec}
\end{equation}
This flux is still subject to the equation of motion~(\ref{feom}) and the
Bianchi identity~(\ref{Bianchi}) which can be written as
\begin{eqnarray}
 d\left[e^{4A}\star\hat{F}\right] &=&0 ,\label{feom1}\\
 d\hat{F} &=& 0\label{Bianchi1}\; .
\end{eqnarray}
Note that the non-linear $F\wedge F$ term in the equation of motion~(\ref{feom})
vanishes in the case at hand since there is no external flux as a result of $f=0$.
Multiplying Eq.~(\ref{diff4b}) with $e^{3A}$, taking the exterior derivative
and using the equation of motion~(\ref{feom1}) one obtains a relation
which differs from Eq.~(\ref{diff5b}) only by terms which involve $d\L_2$.
In fact, it is easy to see this is consistent only if
\begin{equation}
 \L_2 = \mbox{const}\; .
\end{equation}
We have, therefore, concluded that the cosmological constant $\L_2$ of the
four-dimensional space is independent of the internal coordinates.

\vspace{0.4cm}

For later reference, it will be useful to re-write the internal supersymmetry
equations~(\ref{diff2b})--(\ref{diff6b}) by solving
for the exterior derivatives of the various forms involved and by eliminating
$c_1$ and $dA$ using the external relations~(\ref{e1}) and (\ref{e2}). 
One finds
\ba
\label{struct2}
\d\chi&=&-\left( 2e^{-A}\L_2+\frac{1}{2}\sin (\chi )Q\right)V+\lambda\tan(\chi)\sigma ,\\
\label{struct3}
\d V&=&\lambda\left(\frac{2}{3}+\tan^2(\chi)\right)\sigma\wedge V ,\\
\label{struct4}
\d J&=&-\frac{1}{\cos(\chi)}\star\hat{F}-\frac{3\Lambda_2}{\cos(\chi)}e^{-A}\Psi_-
     +\lambda\left(\frac{4}{3}+\tan^2(\chi)\right)J\wedge \sigma\nonumber\\
   &&+\left(\L_2\tan (\chi )e^{-A}-\frac{3+\cos^2(\chi )}{6\cos (\chi )}Q\right) J\wedge V,\\
\label{struct5}
\d\Psi_-&=&-\tan(\chi)\star\hat{F}\wedge V+\left( 3\L_2\tan (\chi )e^{-A}-
          \frac{1}{2}\cos (\chi )Q\right)V\wedge\Psi_-+\lambda\sigma\wedge\Psi_-\nonumber\\
        &&+2\lambda\frac{\sin(\chi)}{\cos^2(\chi)}\sigma\wedge J\wedge V,\\
\label{struct6}
\d\Psi_+&=&-\tan(\chi)\hat{F}-\frac{\Lambda_2}{\cos(\chi)}e^{-A}J\wedge J+
          \frac{\lambda}{\cos^2(\chi)}\sigma\wedge\Psi_+
          -\frac{Q}{2\cos (\chi )}V\wedge\Psi_+ .
\ea

The task is now to solve the supersymmetry conditions~(\ref{e1}), (\ref{e2})
and~(\ref{diff2b})-(\ref{diff6b}) with internal flux~(\ref{Fspec}) subject
to the equation of motion~(\ref{feom1}) and Bianchi identity~(\ref{Bianchi}).
We have not attempted to do this in general but we will analyze some special
cases in the subsequent sections.

The above equations show that the presence of the cosmological constant $\L_2$
and the phase $\chi$ introduces significant complication. We re-iterate
that different normalizations of the internal spinors is necessary for
having solutions with non-vanishing $\L_2$ and, hence, with
$\mbox{AdS}_4$ as the external space. Indeed, setting $\chi =n\pi$ for an
integer $n$, the values corresponding to equal norms of the
spinors $\e_{(1)}$ and $\e_{(2)}$, immediately implies from Eq.~(\ref{struct2})
that $\L_2=0$. In this case, our equations reduce to the ones given
in Refs.~\cite{kaste:0303,dall:0311}.
We will see below that solutions with non-trivial phase $\chi$
and non-vanishing $\L_2$ exist, at least locally. Also note that $\L_2=0$
does not seem to imply, from the above equations, that the phase $\chi$ is trivial.
We will later present an explicit (local) solution with vanishing $\L_2$ and 
non-trivial $\chi$ which shows this is indeed the case.

\subsection{General properties for $\L_2\neq 0$}
\label{sec:gp}

In the case of $\mbox{AdS}_4$ as the external space we can derive a number of
additional general properties by combining supersymmetry conditions
with the equation of motion and Bianchi identity for $\hat{F}$.
More specifically, consider the integrability condition for Eq.~(\ref{diff6b}),
derived by multiplying the equation with $e^{2A}$ and taking the exterior
derivative. Taking into account the Bianchi identity~(\ref{Bianchi1})
as well as the relations (\ref{e2}), (\ref{struct2})-(\ref{struct6})
we find
\begin{equation}
 \L_2\left[ 2\hat{F}\wedge V+2\star\hat{F}\wedge J-4\l\tan (\chi )\s\wedge\Psi_+\wedge V
  +QV\wedge J\wedge J-\frac{2\l}{\cos (\chi )}\s\wedge J\wedge J\right] = 0\; .
\end{equation}
Note this relation is trivially satisfied for vanishing cosmological constant, $\L_2=0$.
However, for $\L_2\neq 0$ and after inserting the explicit expression~(\ref{Fspec})
for $\hat{F}$ and its dual~(\ref{starF}) we obtain a non-trivial constraint on
the internal flux. It turns out, this constraint is satisfied if and only if
\begin{equation}
 W=0\; .
\end{equation}
Note, from Eq.~(\ref{sigma}), this also implies that $\s =0$. We have hence shown
that a non-zero cosmological constant implies the absence of the vector part of
the internal flux. This simplifies Eq.~(\ref{struct3}) to $\d V=0$. We can, therefore,
introduce a coordinate system ${\bf y}=(z^i,y)$, where $i,j,\dots = 4,\dots ,9$
such that
\begin{equation}
 V=\d y\; ,\label{Vy}
\end{equation}
locally. From Eqs.~(\ref{e1}), (\ref{e2}) and (\ref{struct2}) we then learn
that all scalar quantities are independent of $z^i$ and depend on the single
coordinate $y$ only, that is, we have
\begin{equation}
 A=A(y)\; ,\qquad \chi =\chi (y)\; ,\qquad Q=Q(y)\; ,\qquad c_1=c_1(y)\; .
\end{equation}
The two internal equations~(\ref{e1}) and (\ref{e2}) then read
\begin{eqnarray}
 c_1 &=& -\frac{1}{4}\sin (\chi )Q-\frac{3}{2}e^{-A}\L_2 ,\label{e1b}\\
 A' &=& \frac{1}{6}\cos (\chi )Q \label{e2b}\; ,
\end{eqnarray}
where the prime denotes the derivative with respect to $y$. The external
Eqs.~(\ref{struct2})--(\ref{struct6}) simplify to
\ba
\label{struct1b}
\chi'&=&-\half Q\sin(\chi)-2\Lambda_2e^{-A},\\
\label{struct2b}
\d J&=&-\frac{1}{\cos(\chi)}\star\hat{F}-\frac{3\Lambda_2}{\cos(\chi)}e^{-A}\Psi_-
     +\left(\L_2\tan (\chi )e^{-A}-\frac{3+\cos^2(\chi )}{6\cos (\chi )}Q\right) J\wedge V,\\
\label{struct5b}
\d\Psi_-&=&-\tan(\chi)\star\hat{F}\wedge V+\left( 3\L_2\tan (\chi )e^{-A}-
          \frac{1}{2}\cos (\chi )Q\right)V\wedge\Psi_- ,\\
\label{struct6b}
\d\Psi_+&=&-\tan(\chi)\hat{F}-\frac{\Lambda_2}{\cos(\chi)}e^{-A}J\wedge J
          -\frac{Q}{2\cos (\chi )}V\wedge\Psi_+\; ,
\ea
with $V$ given by Eq.~(\ref{Vy}). It is instructive to look at the components
of these equations with indices in the ${\bf z}$ directions only. We
denote the corresponding exterior derivative in these directions by $\hat{\rm d}$.
Then, with the explicit expressions~(\ref{Fspec}) and (\ref{starF}) for
$\hat{F}$ and its dual (remembering that $c_2=0$ in general and $W=Y=0$
for $\L_2\neq 0$) we find
\ba
\hat{\rm d}J&=&-\frac{1}{\cos(\chi)}\left[(c_1+3\Lambda_2e^{-A})\Psi_-+S\right] ,\\
\hat{\rm d}\Psi_-&=&0 ,\\
\hat{\rm d}\Psi_+&=&-\frac{1}{\cos(\chi)}\left[(\Lambda_2e^{-A}-
                    \frac{1}{6}Q)J\wedge J+J\wedge {\cal A}\right]\; .
\ea
These equations can be interpreted as the differential relations for an $\mbox{SU}(3)$
structure, characterized by $(\Psi ,J)$, for the six-dimensional manifolds obtained
by slicing the seven-dimensional internal manifold at constant $y$. It follows that
\begin{equation}
 \hat{\d}\Psi_-=0\; ,\qquad \hat{\d}J\wedge J=0\; , \label{hf}
\end{equation}
the latter using Eqs.~(\ref{su3r2}) and (\ref{SwJ}). These two relations are
precisely the defining conditions for half-flat manifolds~\cite{salamon}. We,
therefore, conclude that for non-zero cosmological constant, $\L_2\neq 0$, the
six-dimensional slices at constant $y$ of our internal manifold are
half-flat manifolds. Note that the conclusions of this sub-section all followed
from the initial result that $W=0$. Hence, they can even be applied for
vanishing cosmological constant, $\L_2=0$, as long as one imposes
the additional requirement $W=0$.

\section{Solution for $\mbox{SU}(3)$ singlet flux}
\label{sec:sf}

Clearly, solving the supersymmetry conditions for general flux is quite difficult.
In this section, we therefore specialize to $\mbox{SU}(3)$ singlet flux
and consider the solutions in this case, both for vanishing and non-vanishing cosmological
constant $\L_2$. To do this, we have to set the non-singlet flux components
${\cal A}$ and $U$ to zero. For $\L_2\neq 0$ we have seen previously that
$W=0$ automatically, while we impose this as an additional condition for $\L_2=0$. 
In both cases the flux is given by
\begin{equation}
 \hat{F} = -\frac{1}{6}QJ\wedge J-c_1\Psi_+\wedge V\; ,\label{Fsing}
\end{equation}
that is, it only contains the two singlet components $Q$ and $c_1$. Together
with the warp factor $A$ and the phase $\chi$ these form a set of four
scalar quantities which, as we will see, are determined by a closed set
of first-order differential equations. We first note, that from the
external supersymmetry condition~(\ref{e1}) $c_1$ is actually fixed
in terms of the other three scalar by
\begin{equation}
 c_1 = -\frac{1}{4}\sin (\chi )Q-\frac{3}{2}e^{-A}\L_2\label{e1c}\; .
\end{equation}
We have mentioned earlier that the results of Section~\ref{sec:gp} can
be applied as long as $W=0$. In particular, we can split internal
coordinates as $(y^m)=(z^i,y)$, where $i,j,\dots = 0,\dots ,9$
and write the vector $V$ as
\begin{equation}
 V=\d y, \label{Vy1}
\end{equation}
locally. The scalars $Q$, $c_1$, $A$ and $\chi$ are then functions
of the single coordinate $y$ only. Moreover, with Eqs.~(\ref{e2b}) and (\ref{struct1b}),
we immediately have two first order differential equations
for the three scalar $A$, $Q$ and $\chi$. A third equation can be
obtained by inserting the flux~(\ref{Fsing}) into the equation
of motion~(\ref{feom1}) or the Bianchi identity~(\ref{Bianchi1})
using the relation~(\ref{struct2})--(\ref{struct6}) to remove
exterior derivatives. Both calculations actually lead to the same result,
a first-order differential equation for $Q$. The set of first-order
differential equations which determines $A$, $Q$ and $\chi$ then reads
\ba
\label{scb1}
A'&=&\frac{1}{6}Q\cos(\chi),\\
\label{scb2}
Q'&=&\frac{1}{12\cos(\chi)}\left[(7+\cos^2(\chi))Q^2
     -24\sin(\chi)\Lambda_2e^{-A}Q-108(\Lambda_2e^{-A})^2\right],\\
\label{scb3}
\chi'&=&-\frac{1}{2}Q\sin(\chi)-2\Lambda_2e^{-A}\; .
\ea
With these three equations and the solutions~(\ref{Vy1}), (\ref{e1c}) for $V$
and $c_1$ we have satisfied the $F$ equation of motion, the Bianchi
identity and all but three supersymmetry relations. These three remaining
equations are the differential relations~(\ref{struct4})--(\ref{struct6})
for $J$ and $\Psi_\pm$. They can be cast into the simple form
\ba
e^{-2\alpha}{\rm d}\left(e^{2\alpha}J\right)&=&3h\Psi_-\label{d1}\\
e^{-3\alpha}{\rm d}\left(e^{3\alpha}\Psi_-\right)&=&0\\
e^{-3\alpha}{\rm d}\left(e^{3\alpha}\Psi_+\right)&=&2hJ\wedge J\label{d3}
\ea
where we have introduced the two functions
\ba
 h&=&\frac{1}{3\cos(\chi)}\left[\frac{1}{4}Q\sin(\chi)-\frac{3}{2}\Lambda_2e^{-A}\right]\label{h}\\
 g&=&\frac{1}{2\cos(\chi)}\left[\frac{1}{6}Q(1+\cos^2(\chi))-\Lambda_2e^{-A}\sin(\chi)\right]
\ea
and $\a$ is defined by
\begin{equation}
 \a ' = g\; .
\end{equation}
As before, the prime denotes the derivative with respect to $y$. It can be verified
by straightforward computation, using the differential equations~(\ref{scb1})--(\ref{scb3}),
that the functions $h$ and $g$ satisfy $h'=hg$. This implies that
\begin{equation}
 he^{-\a}=c\; , \label{hc}
\end{equation}
where $c$ is an arbitrary real constant. The general solution to
Eqs.~(\ref{d1})--(\ref{d3}) then takes the form
\begin{eqnarray}
 J &=& e^{-2\a}\o \label{odef}\\
 \Psi_\pm &=& e^{-3\a}\O_\pm\; ,
\end{eqnarray}
where the forms $\o$ and $\O_\pm$ are $y$-independent and satisfy
the differential relations
\begin{eqnarray}
 \hat{\rm d}w &=& 3c\O_- ,\label{nK1}\\
 \hat{\rm d}\O_- &=& 0,\\
 \hat{\rm d}\O_+ &=& 2c\o\wedge\o\label{nK3}\; .
\end{eqnarray}
We recall that $\hat{\rm d}$ denotes the exterior derivative with 
respect to the six-dimensional coordinates $z^i$. The forms
$\o$ and $\O_\pm$ define a six-dimensional $\mbox{SU}(3)$
structure with respect to a rescaled vielbein $\tilde{e}^i=\exp (\a )e^i$.
Their above differential relations are precisely the ones for
so-called nearly-K\"ahler manifolds~\cite{salamon}. The metric 
corresponding to these solutions therefore takes the form
\ba
{\rm d}s^2&=&e^{2A}{\rm d}s^2_4+e^{-2\a}{\rm d}s^2_{({\rm nK})}+{\rm d}y^2,
\ea
where ${\rm d}s^2_{({\rm nK})}$ is the metric a a nearly-K\"ahler manifold.
The associated flux is given by
\begin{equation}
 F = -\frac{1}{6}Qe^{-4\a}\o\wedge\o - c_1e^{-3\a}\O_+\wedge dy\; .
\end{equation} 
The scalar $c_1$ is determined by Eq.~(\ref{e1c}) and $A$, $Q$ and
$\chi$ have to be determined by solving the differential
equations~(\ref{scb1})--(\ref{scb3}). For $c\neq 0$ the scale
factor $\a$ is given by
\begin{equation}
 e^{-\a}=\frac{c}{h}\; ,
\end{equation}
and is, therefore, determined once a solution for $A$, $Q$ and $\chi$
has been found. Unfortunately, we did not find the general analytic
solution of the system~(\ref{scb1})--(\ref{scb3}) and, given the
complication of the equations, this may well be impossible. We have
numerically integrated this system for a variety of initial conditions.
All numerical solutions we have found exist on a finite interval
in $y$ only, with (curvature) singularities developing at the
interval endpoints. This does not exclude the existence of special solutions,
obtained from particular choices for the initial conditions in
(\ref{scb1})--(\ref{scb3}), which are globally defined for all $y$.
Indeed, the case $c=0$ which we discuss in detail in the next
subsection leads to a special solution which exists on an half-infinite
range in $y$. Unfortunately, we do not know any other special solutions to
(\ref{scb1})--(\ref{scb3}) which are globally defined.

\subsection{A Calabi-Yau solution}
\label{sec:CY}

The case $c=0$ is special in that, from Eq.~(\ref{hc}), it forces the
function h in Eq.~(\ref{h}) to vanish identically. This leads to an additional
algebraic constraint between $A$, $Q$ and $\chi$. It is easily seen that
the solutions to (\ref{scb1})--(\ref{scb3}) in this case are given by
\begin{eqnarray}
 \sin (\chi ) &=& \frac{\L_2}{k}e^{-5A} \label{chisol}\\
 Q &=& 6ke^{4A}\label{QCY}\\
 \a &=& A+\mbox{const}\; ,
\end{eqnarray}
where $k$ is a constant. The warp factor $A$ is determined by the
differential equation
\ba
A'&=&\sqrt{k^2e^{8A}-(\Lambda_2)^2 e^{-2A}}\; .
\ea
The solution to this equation is only real for
$A>A_{\rm crit}=\frac{1}{5}\ln\left(\frac{\Lambda_2}{k}\right)$
and $A$ starts at \mbox{$A(y\rightarrow -\infty)= A_{\rm crit}$}. 
It then increases until it diverges at some finite $y$. One also finds
from Eq.~(\ref{chisol}) that $\sin (\chi )$ 
is unity at $y\rightarrow -\infty$
and evolves to zero at the singularity. Hence, this solution exists on
a half-infinite range in $y$, as previously claimed.

For $c=0$ the differential relations~(\ref{nK1})--(\ref{nK3}) for
nearly-K\"ahler manifolds reduce to the ones for Calabi-Yau three-folds.
Hence the metric for this solution is given by
\begin{equation}
 {\rm d}s^2_{11}=e^{2A}{\rm d}s^2_4+e^{-2A}{\rm d}s^2_{({\rm CY})}+{\rm d}y^2\; ,
\end{equation}
where ${\rm d}s^2_{({\rm CY})}$ is the Ricci-flat metric of a Calabi-Yau three-fold.
The associated flux takes the form
\begin{equation}
 \hat{F}=-k \o\wedge \o+3\Lambda_2e^{-4A}\O_+\wedge {\rm d}y
\end{equation}
As a consistency check, we have verified that this metric and flux indeed
satisfy the Einstein equations~(\ref{geom}).

\subsection{A special solution for vanishing cosmological constant}
\label{sec:L0}

We have previously seen that setting the norms of the two internal
spinors $\e_{(1)}$ and $\e_{(2)}$ equal, or, equivalently, having
a trivial phase $\chi = n\pi$ for $n$ integer, implies a vanishing
cosmological constant. In this subsection we will show the opposite
is not true, that is, examples with vanishing cosmological constant
and non-trivial phase $\chi$ exist. 

To find such an example, we consider the differential
equations~(\ref{scb1})--(\ref{scb3}) for $\L_2=0$. They can then
be written in the simple form
\begin{eqnarray}
 A'&=&\frac{1}{6}XQ\\
 X'&=&\frac{1}{2}(1-X^2)Q\label{Xp}\\
 Q'&=&\frac{7+X^2}{12X}Q^2
\end{eqnarray}
where we have defined $X=\cos (\chi )$. This leads to the implicit solution
\begin{eqnarray}
 A&=&A_0-\frac{1}{6}\ln |1-X^2|\\
 Q&=&K\left|\frac{X^7}{(1-X^2)^4}\right|^{1/6}\label{QL0}\\
 e^{-2\a} &=& \frac{144c^2}{K^2}\left|\frac{1-X^2}{X}\right|^{1/3}\label{alpha}\; ,
\end{eqnarray}
where $A_0$ and $K$ are constants. Finally, $X$ as a function of $y$ can
be obtained by integrating Eq.~(\ref{Xp}). This integral shows the solutions exist
on a half-infinite range where $X$ varies from zero at $y\rightarrow\pm\infty$
to $X=\pm 1$ at some finite value of $y$. At this point the solution
develops a curvature singularity due to the vanishing of the scale factor~(\ref{alpha}).

\section{Embedding into Horava-Witten theory}
\label{sec:hw}

In this section, we would like to analyze the relation between the
$\mbox{SU}(3)$ structure formalism and Horava-Witten
theory~\cite{horava:9510,horava:9603}, that is, M-theory on the
orbifold $\mbox{S}^1/\mbox{Z}_2$. The action for this theory can be
constructed~\cite{horava:9603} by coupling 11-dimensional supergravity
to two 10-dimensional ${\rm E}_8$ gauge multiplets which are located
on the two 10-dimensional fixed planes of the orbifold. The bulk
equations for this theory are identical to the ones for 11-dimensional
supergravity and it is, therefore, reasonable to ask whether any of
the solutions obtained so far can be embedded in Horava-Witten
theory. This amounts to checking whether any of our solutions can be
made to satisfy the boundary conditions of Horava-Witten theory
imposed at the two 10-dimensional fixed planes. In fact, such an
embedding may be quite useful to cut off curvature singularities
at finite proper distance which, as we have seen, can arise in the
M-theory solutions.

The most relevant boundary condition for our purposes is the one on
the spinor $\eta$ which parameterizes supersymmetry
transformations. Let us take the coordinate in the orbifold
direction to be $x^\natural$. In order to ensure, the various
components of the gravitino remain chiral on the boundaries, $\eta$
should satisfy the conditions
\begin{eqnarray}
 \eta &=& \G_\natural\eta ,\label{b1}\\
 \eta ' &=& -\G_\natural \eta '\label{b2}
\end{eqnarray}
at both boundaries, where the prime denotes the derivative with respect to
$x^\natural$. These conditions have to be analyzed for our spinor Ansatz~(\ref{spinorAnsatz})
and we will do this subsequently for solutions with singlet flux.

\subsection{Embedding of solutions with singlet flux?}

We now specialize to the solutions with singlet flux found in Section~\ref{sec:sf}.
For these solutions one of the seven internal coordinates, previously called $y$,
is singled out and the vector $V$ is given by $V={\rm d}y$. It seems natural to identify
the coordinate $y$ with the orbifold direction, that is $x^\natural =y$.
From Eq.~(\ref{structure1}) the spinor Ansatz~(\ref{spinorAnsatz}) then takes the form
\begin{equation}
 \eta = |\e_{(1)}|\left(\theta\otimes\hat{\e}_{(1)}-\l\g\th\otimes (\g_\natural )\hat{\e}_{(1)}\right)\; ,
 \label{spinorspec}
\end{equation}
where $\hat{\e}_{(1)}$ is a normalized version of the spinor $\e_{(1)}$. From Eq.~(\ref{biXib2})
and (\ref{odef}) it follows that the two-form $\o$ on the nearly-K\"ahler manifold
can be written as
\begin{equation}
 \o_{ij} = -i\hat{\e}_{(1)}^T\tilde{\g}_{ij}\hat{\e}_{(1)}\; ,
\end{equation}
where $\tilde{\g}_i$ are the ``curved'' six-dimensional gamma matrices with respect to
the nearly-K\"ahler metric. Given that both this metric and $\o$ are
$y$-independent we conclude that the same it true for the normalized
spinor $\hat{\e}_{(1)}$. This allows us to perform the $y$-derivative
of $\eta$ which is required for the second boundary
condition~(\ref{b2}). Using the spinor normalization~(\ref{Xi}) and (\ref{Xit})
as well as various relations from Section~\ref{sec:sf} it is then
straightforward to work out the boundary conditions. We find that
the two conditions~(\ref{b1}) and (\ref{b2}) are satisfied for
singlet-flux solutions precisely if
\begin{equation}
 \chi = (2n+1)\pi\; ,\qquad A' = 0\; ,\label{bc}
\end{equation}
at both boundaries for integers $n$. From the differential equation~(\ref{scb1})
for $A'$ the second condition is equivalent to $Q=0$ at the boundaries. We can
then numerically integrate the equations~(\ref{scb1})--(\ref{scb3})
starting from one of the boundaries, located at, say, $y=0$, using the initial
conditions $\chi (0)=(2n+1)\pi$ and $Q(0)=0$. This guarantees the conditions~(\ref{bc})
are satisfied at the first boundary. One would then hope that $\chi$ equals the
same or another odd multiple of $\pi$ at some value $y>0$ which could be identified
with the second boundary. The remaining free initial condition $A(0)$ may then
be used to set $Q=0$ at the second boundary. Numerical integration shows that
$\chi$ either returns to the same odd multiple of $\pi$ or approaches one of the
neighbouring even multiples, depending on the choice of $A(0)$, but in both
cases the solution terminates with a singularity. This makes it impossible to
satisfy the conditions~(\ref{bc}) at the second boundary. Our conclusion is,
therefore, that none of the singlet flux solutions can be embedded in Horava-Witten
theory. 
This statement can also be explicitly verified for the two special classes of solutions
which we have found. Consider first the Calabi-Yau solutions of Section~\ref{sec:CY}.
The relation~(\ref{QCY}) implies that $A\rightarrow -\infty$ whenever $Q\rightarrow 0$
and, hence, a curvature singularity. This precludes the implementation of the
boundary condition~(\ref{bc}). 
For the solutions with vanishing $\L_2$, described in Section~\ref{sec:L0}, we
should have $X=\cos (\chi )=-1$ at the boundary. From Eq.~(\ref{QL0}) this
means that $Q\rightarrow\infty$ which contradicts the boundary condition $Q=0$.

\subsection{Perturbative solutions}

Given the complexity of the supersymmetry conditions it is clearly
worthwhile asking about the existence of suitable approximation
schemes and corresponding approximate solutions. An obvious such
scheme is to start with a general solution for vanishing flux, that
is, a direct product of four-dimensional Minkowski space and a
Calabi-Yau space times $S^1$ as the internal space and compute
first-order corrections due to a small but non-vanishing flux to this
solution. In fact, a class of such approximate solutions has been
found in Ref.~\cite{Witten:1996mz,lukas:9710} in the context
Horava-Witten theory. We would now like to analyze how these solutions
fit into the present context of $\mbox{SU}(3)$ structure.

\vspace{0.4cm}

We start be reviewing the solutions in Ref.~\cite{Witten:1996mz,lukas:9710}.
The metric for these solutions has the form
\begin{equation}
 ds^2_{(11)}=(1+b){\rm d}x^\m {\rm d}x^\n\eta_{\m\n}+(g_{ij}+h_{ij}){\rm d}z^i{\rm d}z^j+(1+\r ){\rm d}
             y^2\; .
 \label{wmetric}
\end{equation}
Here $g_{ij}$, where $i,j,\dots = 4,\dots ,9$ is the Ricci-flat metric
on a Calabi-Yau three-fold with K\"ahler form
$\o_{a\bar{b}}=-ig_{a\bar{b}}$ ($a,b,\dots $ and
$\bar{a},\bar{b},\dots$ denote holomorphic and anti-holomorphic
indices on the Calabi-Yau space) and $y=x^\natural$.  Further, $b$, $\r$ and
$h_{ij}$ depend on the internal coordinates $y^m$ and are the
first-order corrections to be determined. They can be expressed in
terms of a $(1,1)$ form $B_{(2)}$ on the Calabi-Yau
space satisfying the gauge condition
\begin{equation}
 \hat{\rm d}\hat{\star}B_{(2)}=0\; .\label{Bp1}
\end{equation}
The star $\hat{\star}$ denotes the six-dimensional Hodge dual with
respect to the Calabi-Yau metric $g_{ij}$ and $\hat{\rm d}$ is the
exterior derivative with respect to the $z^i$ coordinates.  This
two-form is harmonic with respect to the seven-dimensional internal
space, that is, it satisfies
\begin{equation}
 \triangle B_{(2)}+{B_{(2)}}^{\prime\prime} = 0\; ,\label{Bp2}
\end{equation}
where $\triangle$ is the Laplacian on the Calabi-Yau space. In terms
of this $(1,1)$ form the metric corrections can then
be written as~\footnote{Note these relations and subsequent ones differ
from the ones in Ref.~\cite{lukas:9710} by a factor $\sqrt{2}$ due
to a different normalization of the four-form $F$. For simplicity,
we have also dropped various integration constants which can be absorbed
into coordinate redefinitions and are irrelevant for our purposes.} 
\ba
\label{hcorr}
h_{a\bar{b}}&=&i\left(B_{a\bar{b}}-\frac{1}{3}\o_{a\bar{b}}B\right),\\
\label{b}
b&=&\frac{1}{6}B,\\
\label{g}
\r&=&-\frac{1}{3}B \; ,
\ea
where
\begin{equation}
 B=2\o\lrcorner B_{(2)}\; .
\end{equation}
The associated flux can be written as
\begin{equation}
 \star\hat{F} = {\rm d}B_{(2)} = \hat{\rm d}B_{(2)}+B_{(2)}'\wedge dy\; ,\label{Fd}
\end{equation}   
where $\star$ refers to the seven-dimensional Hodge dual and the prime
denotes the derivative with respect to $y$. The corresponding spinor
in the supersymmetry transformation reads
\begin{equation}
 \z_+=e^{-\phi}\z\label{corrspinor}
\end{equation}
where $\z$ is the covariantly constant complex spinor on the Calabi-Yau
space satisfying
\begin{equation}
 \g_{\bar a}\z = 0\; ,\qquad \g_\natural\z = \z\; ,\qquad \z^\dagger\z =1\; .\label{CYspinor}
\end{equation}
The scalar $\phi$ encodes the first-order correction to
the spinor and is explicitly given by
\begin{equation}
 \phi =-\frac{1}{24}B\; .
\end{equation}
Our notation suggests that the spinor in Eq.~(\ref{corrspinor}) should be identified
with the complex spinor $\z_+$ which appeared in the general formalism. That this is indeed
sensible can be seen from the general projection conditions~(\ref{projections})
which imply the relations~(\ref{CYspinor}). The completes the review of the solutions.

\vspace{0.4cm}

We would now like to check that these solutions indeed satisfy the
general supersymmetry conditions summarized in Section~\ref{sec:summary}.
This provides a useful consistency check for the general results in this
paper as well as explicit (approximate) expressions for the forms
$V$, $J$ and $\Psi$ defining the $\mbox{SU}(3)$ structure. We first
remark that the flux~(\ref{Fd}) clearly satisfies the $F$ equations
of motion~(\ref{feom1}) and Bianchi identity~(\ref{Bianchi1})
to leading order as a result of the properties~(\ref{Bp1}) and (\ref{Bp2}).

Next we should identify the quantities which enter the general formalism
in terms of the approximate solutions. Note first that we are working with
four-dimensional Minkowski space and, hence,
\begin{equation}
 \L_2=0\; .
\end{equation}
Further, from Eqs.~(\ref{corrspinor}) and (\ref{CYspinor}) we have
$0=\z_+^T\g_{a\bar{b}}\z_+= 2g_{a\bar{b}}\z_+^T\z_+$. Using
Eq.~(\ref{biXiTilde}), we conclude that $\tilde{\Xi}=0$ which implies,
by virtue of Eq.~(\ref{Xit}), that
\begin{equation}
\sin (\chi )=0\; .
\end{equation}
We are, therefore, dealing with solutions where the two norms of the
Majorana spinors are identical. By comparing the metric Ans\"atze~(\ref{wmetric})
and (\ref{11metric}) we identify the warp factor to leading order as
\begin{equation}
 A=\frac{b}{2}=\frac{B}{12}\; . \label{A}
\end{equation}
For the singlet and vector components of the flux one finds
\begin{eqnarray}
 Q &=& -\frac{1}{2}B' ,\\
 c_1 &=& 0 ,\\
 \s &=& \frac{1}{4}\hat{\rm d}B\; .
\end{eqnarray}
Finally, one has to compute the approximate $\mbox{SU}(3)$ structure
$(V,J,\Psi )$ which can be done using the bi-spinor expressions in
Appendix~\ref{AppB} and inserting the spinor~(\ref{corrspinor}) and
the internal metric from Eqs.~(\ref{wmetric}) and (\ref{hcorr}).
The result to first order in the flux is
\begin{eqnarray}
 V&=&\left(1-\frac{B}{6}\right) {\rm d}y ,\label{Va}\\
 J&=&\o+B_{(2)}-\frac{1}{3}B\o ,\\
 \Psi &=&\O-\frac{1}{4}B\O , \label{Psia}
\end{eqnarray}
where $\O$ is the $(3,0)$ form on the Calabi-Yau space. 
With these identifications it is straightforward to verify that
the supersymmetry conditions~(\ref{e1}), (\ref{e2}) and
(\ref{diff2b})--(\ref{diff6b}) are indeed satisfied to first order in
the flux, provided we choose the phase $\chi$ so that $\cos (\chi )=-1$.

\vspace{0.4cm}

Are these the only perturbative supersymmetric solutions obtained from
zeroth order solutions which are direct products of four-dimensional
Minkowski space, a circle and a Calabi-Yau three-fold? The answer is
yes as long as one requires that the cosmological constant remains
zero under the perturbations~\footnote{This means that the topology of
the four-dimensional space should not change when including the
first-order corrections, an assumption which seems reasonable in a
perturbative calculation.}.  To verify this statement, first note that
from Eq.~(\ref{e1}) we have $c_1=0$ since $\L_2=0$ and $\sin (\chi
)=0$ to zeroth order.  This means that~(\ref{Fd}) represents the most
general flux where $B_{(2)}$ is a $(1,1)$ form. One can insert this
flux into Eqs.~(\ref{e2}), (\ref{diff3b})--(\ref{diff6b}) and
integrate these equations to obtain precisely the scale
factor~(\ref{A}) and the expressions~(\ref{Va})--(\ref{Psia}) for the
$\mbox{SU}(3)$ structure.  The warped Calabi-Yau solutions of
Ref.~\cite{Witten:1996mz,lukas:9710} are therefore the only
perturbative solutions of Horava-Witten theory starting from the
direct product of Calabi-Yau space, circle and Minkowski space at
lowest order. It is not clear to us whether or not there exist other
solutions of Horava-Witten, corresponding to $N=1$ four-dimensional
supersymmetry, which cannot be obtained in this perturbative
manner. We have shown for the singlet-flux solutions of
Section~\ref{sec:sf} that an embedding into Horava-Witten theory is
not possible. However, there may still be other solutions with a more
complicated structure of the flux which do allow for such an embedding.

\section{Conclusion}

In this paper, we have analyzed supersymmetric solutions of M-theory
based on a seven-dimensional internal space with $\mbox{SU}(3)$
structure and a four-dimensional maximally symmetric space, either
four-dimensional Minkowski space or $\mbox{AdS}_4$. We have worked
with what we believe is the most general Ansatz for the fields and
particularly the supersymmetry spinor and have derived the complete
set of supersymmetry conditions in this case. These conditions are
summarized in Section~\ref{sec:summary}. They are more general and
more complicated than related equations previously derived in the
literature since we have allowed the norm of the two spinors which
define the $\mbox{SU}(3)$ structure to be different. This introduces a
new degree of freedom into the problem which can be conveniently
parameterized by a phase $\chi$ and which has been ignored in earlier
work, leading to incorrect conclusions.
When $\chi$ is set to one of its trivial values $\chi = n\pi$,
corresponding to equal norms of the two spinors, the equations of
Section~\ref{sec:summary} reduce to the ones previously
derived~\cite{kaste:0303,dall:0311}.

\vspace{0.4cm}

We have obtained a number of general results which must hold for all
supersymmetric M-theory solutions of the type considered in this
paper. Specifically, we have found that the external flux in the
direction of the four-dimensional maximally symmetric space and $c_2$,
one of the three singlets contained in the internal flux, always
vanish. Further, we have seen that from the two types of Killing
spinors on $\mbox{AdS}_4$ (see Eq.~\ref{4Dspinor}) only one type can be
compatible with supersymmetry and that the associated cosmological
constant $\L_2$ must be independent of the internal coordinates. It
also turned out that having different norms of the two spinors, or
equivalently, having a nontrivial phase $\chi\neq n\pi$, is a
necessary condition for the existence of solutions with non-zero
cosmological constant and, hence, $\mbox{AdS}_4$ as the external
space. 

For non-vanishing cosmological constant, $\L_2\neq 0$, we have
arrived at a number of additional general conclusions. We found that
in this case the vector part $W$ of the internal flux always
vanishes. This implied the vector $V$, which together with
the two-form $J$ and the complex three-form $\Psi$ defines
the seven-dimensional $\mbox{SU}(3)$ structure, could be written
as $V={\rm d}y$ for some internal coordinate $y$. Moreover, all
singlet quantities, namely the two singlet flux components
$Q$ and $c_1$ as well as the warp factor $A$ and the phase
$\chi$, turned out to be functions of $y$ only. We also
showed that the six-dimensional spaces, obtained by slicing
the internal space at constant $y$, are always half-flat
manifolds.   

\vspace{0.4cm}

As an application of our results, we have considered the case of
singlet flux where only the two $\mbox{SU}(3)$ singlets $Q$ and $c_1$
of the internal flux are non-vanishing. We showed the internal
seven-dimensional space is a warped product of a circle with
coordinate $y$ and a six-dimensional nearly-K\"ahler manifold, in this
case. The problem of finding explicit solutions has been reduced to
the one of finding solutions to a complicated system of three
first-order differential equations for the warp factor $A$, the
singlet flux $Q$ and the phase $\chi$. Special solutions include a
case where the nearly-K\"ahler manifold degenerates to a Calabi-Yau
three-fold but $\L_2\neq 0$ and another case with $\L_2=0$, a
nearly-K\"ahler manifold and a non-trivial phase $\chi$. The latter
example shows that a vanishing cosmological constant does not imply a
trivial phase $\chi$. Hence, even for Minkowski space as the external
space are our equations more general than the ones obtained for
equal spinor norms. We did not find the general analytic
solution for the first-order differential equations which determine
$A$, $Q$ and $\chi$. Numerically solutions exist on a finite interval
in $y$ with the interval endpoints corresponding to curvature
singularities. This, however, does not exclude the existence of
globally defined solutions which may arise from special initial
values. Indeed, the two special solutions mentioned above exist on a
half-infinite range in $y$ rather than an interval. It would be
interesting to study the global properties of the differential
equations~(\ref{scb1})--(\ref{scb3}) to see whether solutions defined
for all $y$ do or do not exist.

\vspace{0.4cm}

Finally, we have discussed the embedding of solutions into
Horava-Witten theory. This problem is of some phenomenological
relevance since such solutions would lead to heterotic
vacua with $N=1$ supersymmetry in four dimensions. Unfortunately, we
found none of the singlet-flux solutions could be made to satisfy the
appropriate heterotic boundary conditions. As a consistency check we
have also verified that the approximate warped Calabi-Yau solutions of
Horava-Witten theory described in Ref.~\cite{Witten:1996mz,lukas:9710}
satisfy our general equations. We have also shown they are the only
perturbative solutions of this theory based on a lowest-order space
which is a direct product of four-dimensional Minkowski space, a
Calabi-Yau three-fold and a circle. It would be interesting to see
whether there exist any exact solutions for a more complicated
structure of the flux which satisfy heterotic boundary conditions.

\vspace{1cm}
\noindent
{\large\bf Acknowledgments} We would like to thank James Sparks and
Daniel Waldram for discussions. Both authors are supported by
a PPARC Advanced Fellowship.\\


\vskip 1cm
\appendix{\noindent\Large \bf Appendix}
\renewcommand{\theequation}{\Alph{section}.\arabic{equation}}
\setcounter{equation}{0}


\section{Gamma matrices and spinors in $d=4,7,11$}
\label{AppA}

In this section, we summarize our conventions for gamma matrices and
spinors in dimensions four, seven and eleven. We begin with four dimensions.

\vspace{0.4cm}

Four-dimensional curved indices are denoted by $\m ,\n ,\dots =
0,1,2,3$ and their flat counterparts by $\muul , \nuul ,\dots =
0,1,2,3$. We adopt the convention that explicit indices refer to
tangent space indices. The gamma matrices $\g_\muul$ are
$4\times 4$ matrices satisfying the standard anti-commutation relations
\begin{equation}
 \{\g_\muul ,\g_\nuul\} = 2\eta_{\muul\nuul}
\end{equation}
where $(\eta_{\muul ,\nuul}) = \mbox{diag}(-1,+1,+1,+1)$ is the ``mostly plus'' Minkowski metric.
Their hermitian conjugate $(\g_\muul )^\dagger$ is given by
\begin{equation}
 (\g_\muul )^\dagger = \g_0\g_\muul\g_0\; .
\end{equation}
We normalize the Levi-Civita tensor $\e_{\muul\nuul\rhul\siul}$ by
$\e_{0123}=1$ and define
\begin{equation}
 \g \equiv \frac{i}{4!}\e_{\muul\nuul\rhul\siul}\g^{\muul\nuul\rhul\siul}
     =-i\g_0\g_1\g_2\g_3\; , \label{g5}
\end{equation}
where a $\g$ with multiple indices denotes the anti-symmetrized product
of gamma matrices, as usual. Clearly, we have $\g^2=\UM_4$. The generators
of the four-dimensional Lorentz group are given by
\begin{equation}
 \s_{\muul\nuul} = \frac{1}{4}[\g_\muul ,\g_\nuul ]=\frac{1}{2}\g_{\muul\nuul}
\end{equation}
For convenience, we will use a real representation of the gamma matrices $\g_\muul$
so that $\g$ in Eq.~(\ref{g5}) is purely imaginary. Then, spinors $\th\in\CM^4$
are Majorana precisely when they are real. Weyl-spinors are characterized by
$\g\th =\pm \th$, as usual.

\vspace{0.4cm}

Let us now turn to seven dimensions. As curved indices we take $m,n,\dots = 4,\dots ,9,\natural$
with flat counterparts $\mul ,\nul ,\dots = 4,\dots ,9,\natural$. As before, explicit indices
refer to tangent space indices. The gamma matrices $\g_\mul$ are hermitian $8\times 8$
matrices satisfying
\begin{equation}
 \{\g_\mul ,\g_\nul\}=2\delta_{\mul\nul}\; .
\end{equation}
The normalization is fixed by the relation
\begin{equation}
 \g_4\dots\g_9\g_\natural = -i\UM_8\; .
\end{equation}
The generators of $\mbox{SO}(7)$ are given by
\begin{equation}
 \s_{\mul\nul}=\frac{1}{4}[\g_\mul ,\g_\nul ]=\frac{1}{2}\g_{\mul\nul}\; .
\end{equation}
We will use a purely imaginary representation of the gamma matrices, $\g_{\mul}^*=-\g_\mul$,
so that
\begin{equation}
 (\g_\mul )^T=-\g_\mul\; ,\qquad (\g_{\mul\nul})^T=-\g_{\mul\mul}\; ,\qquad
 (\g_{\mul\nul\pul})^T=\g_{\mul\nul\pul}\; . \label{t7}
\end{equation}
In this representation, spinors $\z\in\CM^8$ are Majorana precisely when they are real.  
    
\vspace{0.4cm}

Finally, in eleven dimensions curved indices are denoted by $M,N,\dots = 0,\dots ,9,\natural$
with flat counterparts $\Mul ,\Nul ,\dots = 0,\dots ,9,\natural$. Explicit indices refer to
tangent space indices. The gamma matrices are $32\times 32$ matrices $\G_\Mul$ satisfying
\begin{equation}
 \{\G_\Mul ,\G_\Nul\}=2\eta_{\Mul\Nul}\; ,
\end{equation}
where $(\eta_{\Mul\Nul})=(-1,+1,\dots ,+1)$ is the ``mostly-plus'' Minkowski metric.
For the hermitian conjugate we have
\begin{equation}
(\G_\Mul )^\dagger = \G_0\G_\Mul\G_0\; ,
\end{equation}
and the normalization is such that
\begin{equation}
 \G_0\dots\G_9\G_\natural = \UM_{32}\; .
\end{equation}
The generators of the 11-dimensional Lorentz group read
\begin{equation}
 \S_{\Mul\Nul} = \frac{1}{4}[\G_\Mul ,\G_\Nul ] =\frac{1}{2}\G_{\Mul\Nul}\; .
\end{equation}
For convenience we use a real representation of the gamma matrices $\G_\Mul$.
Then, spinors $\eta\in\CM^{32}$ are Majorana if they are real. Such a real
representation can be obtained in terms for the four-dimensional and seven-dimensional
gamma matrices introduced above by setting
\begin{equation}
 \G_{\muul}=\g_\muul\otimes\UM_8\; ,\qquad
 \G_{\mul}=\g\otimes\g_\mul\; .
\end{equation}
Note these matrices are indeed real given that $\g_\muul$ are real and $\g_\mul$
and $\g$ are purely imaginary. 

\section{SU$(3)$ structures in seven dimensions}
\label{AppB}

In this Appendix we discuss the general formalism dealing with
seven-dimensional $\mbox{SU}(3)$ structure
manifolds~\cite{kaste:0303,dall:0311}. These manifolds form the
``internal'' part of the supersymmetric solutions of M-theory
considered in this paper.

\vspace{0.4cm}

Consider a seven-dimensional spin manifold $X$ with two Majorana spinors
$\e_{(i)}$, where $i=1,2$, linear independent at each point. The
existence of these two spinors implies the reduction of the generic
structure group $\mbox{Spin}(7)$ of the spin bundle to
$\mbox{SU}(3)$. The spinors then correspond to the two singlets in the
decomposition of the ${\bf 8}$ representation of $\mbox{Spin}(7)$
under $\mbox{SU}(3)$. It is such seven-dimensional manifolds with two
spinors $\e_{(i)}$ and structure group $\mbox{SU}(3)$ which we are interested in.

Alternatively, one can discuss the reduction of the structure group in
two steps. A single Majorana spinor $\e_{(1)}$ reduces
$\mbox{Spin}(7)$ to a $\mbox{G}_2$ subgroup, implying the existence of
a $\mbox{G}_2$ structure.  This spinor corresponds to the singlet in
the decomposition of the ${\bf 8}$ representation of $\mbox{Spin}(7)$
under $\mbox{G}_2$. A second, independent spinor $\e_{(2)}$ leads
to another $\mbox{G}_2$ subgroup of $\mbox{Spin}(7)$ in the same way. These two
$\mbox{G}_2$ subgroups intersect in $\mbox{SU}(3)$, which is the relevant
structure group in the presence of two independent spinors.
  
Manifolds with $\mbox{SU}(3)$ structure can also be characterized 
by a collection of $\mbox{SU}(3)$-invariant forms which can be obtained as
bi-linears of the spinors $\e_{(i)}$. To this end, it is useful to define
the complex spinors
\begin{equation}
 \z_\pm = \e_{(1)}\pm i\e_{(2)}\; .\label{zeta}
\end{equation}
We can now introduce the following forms
\ba
\label{biXi}
\Xi&=&\zeta_+^\dagger\zeta_+,\\
\label{biXiTilde}
\tilde{\Xi}&=&\zeta_+^\dagger\zeta_-,\\
\label{biXi1}
\Xi_m&=&-\zeta_+^\dagger\gamma_m\zeta_+=\zeta_-^\dagger\gamma_m\zeta_-,\\
\label{biXi2}
\Xi_{mn}&=&i\zeta_+^\dagger\gamma_{mn}\zeta_+=-i\zeta_-^\dagger\gamma_{mn}\zeta_-,\\
\label{biXi3}
\Xi_{mnp}&=&i\zeta_+^\dagger\gamma_{mnp}\zeta_+,\\
\label{biXi3Tilde}
\tilde{\Xi}_{mnp}&=&-\zeta_+^\dagger\gamma_{mnp}\zeta_-\; ,
\ea
where we have used seven-dimensional curved indices~\footnote{In this
Appendix we only deal with seven-dimensional objects and we will,
unlike in the rest of the paper, use the index range $1,\dots ,7$, for
simplicity. If results in this Appendix are used in the general part of the paper,
the indices should be mapped according to $(1,\dots ,7)\rightarrow (4,\dots ,9,\natural)$.}
$m,m,\dots =1,\dots ,7$. Their flat counterparts will be denoted by
$\mul ,\nul ,\dots =1,\dots ,7$ and explicit indices always refer to
tangent space indices. Although it is possible to work with these
spinor bi-linears while leaving $\e_{(i)}$ general, it is much more
useful, following \cite{bryant:0004,hackett:0306,gauntlett:0302}, to
pick a particular basis where the spinors satisfy a series of
projection conditions.  A convenient choice for these conditions is
provided by \ba
\label{projections}
\gamma_{1357}\epsilon_{(1)}&=&-\epsilon_{(1)},\quad\gamma_{1357}\epsilon_{(2)}= \epsilon_{(2)},\\
\gamma_{1256}\epsilon_{(1)}&=&-\epsilon_{(1)},\quad\gamma_{1256}\epsilon_{(2)}=-\epsilon_{(2)},\\
\gamma_{1234}\epsilon_{(1)}&=&-\epsilon_{(1)},\quad\gamma_{1234}\epsilon_{(2)}=-\epsilon_{(2)}\; .
\ea
Note each of the three conditions for a given spinor projects onto a four-dimensional subspace
and, together, they restrict the spinor to a one-dimensional subspace. Further, as we will see, the
conditions are chosen so that the two one-dimensional subspaces for $\e_{(1)}$ and
$\e_{(2)}$ are orthogonal. We specify the frame in which the above conditions hold
by vielbein one-forms $e^\mul$ and also introduce three complex one forms $\x^a$,
where $a,b,\dots =1,2,3$, and their complex conjugates $\x^{\bar{a}}$, where
$\bar{a},\bar{b},\dots =\bar{1},\bar{2},\bar{3}$ by
\begin{eqnarray}
 \x^1 &=& e^1+ie^2\; ,\\
 \x^2 &=& e^3+ie^4\; ,\\
 \x^3 &=& e^5+ie^6\; .
\end{eqnarray}
It is easy to see that the spinor $\g_{12}\e_{(1)}$ satisfies the same projection
conditions as $\e_{(2)}$. Hence, we must have
\ba
\label{norm12}
\epsilon^2&=&\lambda\gamma_{12}\epsilon^1,
\ea
for a real function $\l$. It is clear that $\l$ determines
the relative normalization of the two spinors, that is,
\begin{equation}
 |\e_{(2)}|^2=\l^2|\e_{(1)}|^2 ,\label{norm3}
\end{equation}
with the norm $|\cdot |$ defined by
\begin{equation}
 |\e_{(i)}|^2=\e_{(i)}^T\e_{(i)}\; .
\end{equation}
Another important conclusion from Eq.~(\ref{norm12}) is that
\begin{equation}
\e_{(1)}^T\e_{(2)}=0\; ,
\end{equation}
since, from Eq.~(\ref{t7}), $\g_{12}$ is an anti-symmetric matrix.
This confirms that the two spinors are indeed orthogonal as stated above.

The above projection conditions can be used to systematically express
the forms~(\ref{biXi})--(\ref{biXi3Tilde}) in terms of the vielbein one-forms
$e^\mul$. One finds
\ba
\label{biXib}
\Xi&=&(1+\lambda^2)|\epsilon_{(1)}|^2,\\
\label{biXibtilde}
\tilde{\Xi}&=&(1-\lambda^2)|\epsilon_{(1)}|^2,\\
\label{biXib1}
\Xi_{(1)}&=&2\lambda |\epsilon_{(1)}|^2V,\\
\label{biXib2}
\Xi_{(2)}&=&2\lambda |\epsilon_{(1)}|^2 J,\\
\label{biXib3}
\Xi_{(3)}&=&|\epsilon_{(1)}|^2\left[ (1-\lambda^2)\Psi_-
                       -(1+\lambda^2)J\wedge V\right],\\
\label{biXib3tilde}
\tilde{\Xi}_{(3)}&=&2\lambda |\epsilon_{(1)}|^2\Psi_+
         +i|\epsilon_{(1)}|^2\left[ (1+\lambda^2)\Psi_--(1-\lambda^2)J\wedge V\right]\; ,
\ea
where the forms $V$, $J$ and $\Psi$ are defined as
\ba
\label{su3V}
V&=&e^7,\\
\label{su3J}
J&=&\frac{i}{2}\left(\xi^1\wedge\xi^{\bar{1}}+\xi^2\wedge\xi^{\bar{2}}+
\xi^3\wedge\xi^{\bar{3}}\right),\\
\label{su3psi}
\Psi&=&\Psi_++i\Psi_-=\xi^1\wedge\xi^2\wedge\xi^3\; .
\ea

We have initially set up the seven-dimensional $\mbox{SU}(3)$
structure by postulating the existence of two Majorana spinors
$\e_{(1)}$ and $\e_{(2)}$.  Alternatively, the $\mbox{SU}(3)$
structure can now be characterized by a one-form $V$, a two-form
$J$ and a complex three-form $\Psi$ which, in terms of the vielbein
$e^\mul$ take the form~(\ref{su3V})--(\ref{su3psi}). The relation
between those two approaches of characterizing an
$\mbox{SU}(3)$ structure is, of course, provided by the above
Eqs.~(\ref{biXi})--(\ref{biXi3Tilde}) and
(\ref{biXib})--(\ref{biXib3tilde}). It is easy to see from
(\ref{su3V})--(\ref{su3psi}) that $V$, $J$ and $\Psi$
satisfy
\begin{eqnarray}
 \Psi\wedge\bar{\Psi}&=&-\frac{4i}{3}J\wedge J\wedge J\label{su3r1},\\
 \Psi\wedge J &=& 0,\label{su3r2}\\
 V\lrcorner J &=& 0,\label{su3r3}\\
 V\lrcorner\Psi &=&0 ,\label{su3r4}\\
 V\lrcorner V &=& 1 , \label{VV}\\
 {J^m}_n{J^n}_p &=& -\delta^m_p+V^mV_p, \label{JJ}\\
 {J_m}^n\Psi_{\pm npq} &=& \mp\Psi_{\mp mpq}, \label{Jpsi}\\
 \star\Psi_\pm &=& \pm\Psi_\mp\wedge V,\\
 \star (J\wedge V) &=& \frac{1}{2}J\wedge J\; .\label{JwV}
\end{eqnarray}
Here, $\lrcorner$ denotes the contraction of two forms or, more precisely,
\begin{equation}
(\alpha\lrcorner\beta)_{i_1\dots i_n}=\frac{1}{p!}\alpha^{j_1\dots j_p}
   \beta_{j_1\dots j_pi_1\dots i_n},
\end{equation}
for a $p$-form $\alpha$ and a $(p+n)$-form $\b$.  Note that from
Eq.~(\ref{JJ}), the tensor ${J^m}_n$ can be viewed as an almost
complex structure (suitably generalized to seven dimensions by
including the vector $V$). It can be used to introduce holomorphic and
anti-holomorphic indices in the usual way. Let us refer to a form
with $q$ holomorphic, $r$ anti-holomorphic and $s\in \{ 0,1\}$ singlet
indices (in the direction of $V$) as a $(q,r,s)$ form.  In this
terminology, $V$ is a $(0,0,1)$ form, $J$ is $(1,1,0)$ form and $\Psi$
is a $(3,0,0)$ form, the latter as a result of Eq.~(\ref{Jpsi}).

\vspace{0.4cm}

Let us briefly digress to explain the precise relation between the above
$\mbox{SU}(3)$ structure and the two $\mbox{G}_2$ structures associated
with $\e_{(1)}$ and $\e_{(2)}$. To this end we introduce the two three-forms
\begin{equation}
 \Phi^{(i)}_{mnp}=i\e_{(i)}^T\g_{mnp}\e_{(i)}\label{Phi}\; ,
\end{equation}
where $i=1,2$. Using the projection conditions~(\ref{projections}) it is
straightforward to show that
\begin{equation}
 \Phi^{(i)}=|\e_{(i)}|^2\phi^{(i)},\label{phi}
\end{equation}
where
\begin{eqnarray}
 \phi^{(1)} &=& \Psi_--J\wedge V = \mbox{Im}(\x^1\wedge\x^2\wedge\x^3)-\frac{i}{2}
            (\x^1\wedge\x^{\bar 1}+\x^2\wedge\x^{\bar 2}+\x^3\wedge\x^{\bar 3})\wedge e^7,
  \label{phi1} \\
 \phi^{(2)} &=& -\Psi_--J\wedge V = -\mbox{Im}(\x^1\wedge\x^2\wedge\x^3)-\frac{i}{2}
            (\x^1\wedge\x^{\bar 1}+\x^2\wedge\x^{\bar 2}+\x^3\wedge\x^{\bar 3})\wedge e^7\; .
  \label{phi2}
\end{eqnarray}
The two $\mbox{G}_2$ structures can either be characterized by the two spinors
$\e_{(i)}$ or the two three-forms $\phi^{(i)}$ in Eqs.~(\ref{phi1}) and (\ref{phi2})
and the relation between these two viewpoints is provided by Eqs.~(\ref{Phi}) and
(\ref{phi}). 

\vspace{0.4cm}

When we come to study the conditions that supersymmetry places on the
bi-linears is will be useful to have some of the above equations for
the spinors $\e_{(i)}$ available in a covariant form. For example,
with the help of the projections~(\ref{projections}) it is easy to see that
the relation~(\ref{norm12}) between $\e_{(1)}$ and $\e_{(2)}$ can be
covariantly written as
\begin{equation}
\label{structure1}
 \epsilon^2=i\lambda V^m\gamma_m\epsilon^1\; .
\end{equation}
For the action of gamma matrices on the spinors we find the covariant results
\ba
\label{structure2}
\gamma^{mn}\epsilon_{(1)}&=&i(J\wedge V)^{mnp}\gamma_p\epsilon_{(1)}
                        -i\Psi_-^{mnp}\gamma_p\epsilon_{(1)},\\
\gamma^{mnp}\epsilon_{(1)}&=&i(J\wedge V)^{mnp}\epsilon_{(1)}
                         -i\Psi_-^{mnp}\epsilon_{(1)}
                         +(\Psi_+\wedge V)^{mnpq}\gamma_q\epsilon_{(1)}\nonumber\\
                        && +\half(J\wedge J)^{mnpq}\gamma_q\epsilon_{(1)},\label{structure3}\\
\gamma^{mnpq}\epsilon_{(1)}&=&-\half(J\wedge J)^{mnpq}\epsilon_{(1)}
                        -(\Psi_+\wedge V)^{mnpq}\epsilon_{(1)}
                        -4i(J\wedge V)^{[mnp}\gamma^{q]}\epsilon_{(1)}\nonumber\\
                      && +4i\Psi_-^{[mnp}\gamma^{q]}\epsilon_{(1)},\label{structure4}\\
\label{structure5}
\gamma^{mnpqr}\epsilon_{(1)}&=&-5(\Psi_+\wedge V)^{[mnpq}\gamma^{r]}\epsilon_{(1)}
                         -\frac{5}{2}(J\wedge J)^{[mnpq}\gamma^{r]}\epsilon_{(1)}.
\ea

\vspace{0.4cm}

The $\mbox{SU}(3)$ structure defined by $(\Psi ,J,V)$ can be used to
decompose an anti-symmetric tensor field on the seven-dimensional manifold
into its various ${\rm SU}(3)$ parts. For us, the relevant case is that
of a four-form and we will briefly explain the relevant details~\cite{dall:0311}
in this case. A four-form in seven dimensions transforms in the
${\bf 35}_{{\rm SO}(7)}$ representation of the generic structure group
$\mbox{SO}(7)$. Under $\mbox{SU}(3)$ this representation decomposes as
\begin{equation}
  {\bf 35}_{{\rm SO}(7)} \rightarrow \left[ 3\;{\bf 1}+2\; ({\bf 3}+{\bf\bar{3}})
                            +({\bf 6}+{\bf\bar{6}})+{\bf 8}\right]_{{\rm SU}(3)}\; .
\end{equation}
We denote the three singlets by $Q$, $c_1$ and $c_2$ and the two
vectors in ${\bf 3}+{\bf\bar{3}}$ as $Y$ and $W$. They are
characterized by the constraints
\begin{equation}
 V\lrcorner Y=V\lrcorner W=0\; .
\end{equation}
Both $Y$ and $W$ consist of a $(1,0,0)$ and a $(0,1,0)$ part. Further,
the octet in the above decomposition can then be represented by a
traceless $(1,1,0)$ form which we denote by ${\cal A}$. This two-form
${\cal A}$ can alternatively be characterized by the
coordinate-independent constraints
\begin{equation}
 J\lrcorner{\cal A}=0\; ,\qquad {\cal A}\lrcorner\Psi =0\; ,\qquad V\lrcorner{\cal A}=0\; .
\end{equation}
Likewise, the ${\bf 6}+{\bf\bar{6}}$ part can be represented by a
three-form $U$ which contains a $(2,1,0)$ and a $(1,2,0)$ part and is
characterized by
\begin{equation}
 \Psi\lrcorner U=0\; ,\qquad J\lrcorner U =0\; ,\qquad V\lrcorner U = 0\; .
\end{equation}
For convenience, we have summarized these components and their
properties in Table~\ref{tab:1}.
\begin{table}
\begin{center}
\begin{tabular}{|l|l|l|l|}
 \hline
 Name&$\mbox{SU}(3)$ representation& index type&constraints\\\hline\hline
 $Q$,$c_1$,$c_2$&${\bf 1}$&$(0,0,0)$&none\\\hline
 $Y$&${\bf 3}+{\bf\bar 3}$&$(1,0,0)$, $(0,1,0)$&$V\lrcorner Y=0$\\\hline
 $W$&${\bf 3}+{\bf\bar 3}$&$(1,0,0)$, $(0,1,0)$&$V\lrcorner W=0$\\\hline
 ${\cal A}$&${\bf 8}$&$(1,1,0)$&${\cal A}\lrcorner\Psi =0$\\
 &&&$J\lrcorner{\cal A}=0$\\
 &&&$V\lrcorner{\cal A}=0$\\\hline
 $U$&${\bf 6}+{\bf\bar 6}$&$(2,1,0)$, $(1,2,0)$&$\Psi\lrcorner U =0$\\
 &&&$J\lrcorner U=0$\\
 &&&$V\lrcorner U=0$\\\hline
\end{tabular}
\caption{Various $\mbox{SU}(3)$ components of a four-form field on a
         seven-dimensional manifold.}
\label{tab:1}
\end{center}
\end{table}
A general four-form $\hat{F}$ can now be expressed as
\begin{equation}
\label{formAnsatz}
\hat{F}=-\frac{1}{6}QJ\wedge J+J\wedge {\cal A}+\Psi_-\wedge Y
           -c_1\Psi_+\wedge V-c_2\Psi_-\wedge V+V\wedge J\wedge W+V\wedge U\; .\\
\end{equation}
Using the explicit expressions~(\ref{su3V})--(\ref{su3psi}) for $V$, $J$ and
$\Psi$ along with the constraints listed in Table~\ref{tab:1} one obtains
the Hodge-dual
\begin{equation} 
\label{starF}
\star \hat{F}=-\frac{1}{3}QJ\wedge V-{\cal A}\wedge V-V\wedge (Y\lrcorner \Psi_+)
           +c_1\Psi_--c_2\Psi_+
           +J\wedge (W\lrcorner J)+S\; ,
\end{equation}
where $S$ is a three-form defined by
\begin{equation}
 S = \star (V\wedge U)\; .
\end{equation}
It contains a $(2,1,0)$ and a $(1,2,0)$ part and can be characterized as
a three-form satisfying the constraints
\begin{equation}
 \Psi\lrcorner S =0\; ,\qquad J\lrcorner S =0\; ,\qquad V\lrcorner S =0\; .
\end{equation}
The second of these realtions implies that $S\lrcorner (J\wedge J)=0$ and using Eq.~(\ref{JwV})
we conclude that
\begin{equation}
 S\wedge J=0\; .\label{SwJ}
\end{equation}



\begin{thebibliography}{10}

\bibitem{gibbons:82}
G. Gibbons and C. Hull,
\newblock ``A Bogomolny bound for general relativity and solitons in
{\cal N}=2 supergravity'',
\newblock Phys.\ Lett.\ B {\bf 109} (1982) 190.

\bibitem{tod:83}
K. Tod,
\newblock ``All metrics admitting supercovariantly constant spinors'',
\newblock Phys.\ Lett.\ B {\bf 121} (1983) 241.

\bibitem{candelas:84}
P. Candelas and D. Raine,
\newblock ``Spontaneous compactification and supersymmetry in d=11 supergravity''.
\newblock Nucl.\ Phys.\ B {\bf 248} (1984) 415.

\bibitem{gauntlett:0205}
J. Gauntlett, D. Martelli, S. Pakis and D. Waldram,
\newblock ``G-structures and wrapped NS5-branes''.
\newblock hep-th/0205050.

\bibitem{gauntlett:0209}
J. Gauntlett, J. Gutowski, C. Hull, S. Pakis and H. Reall,
\newblock ``All supersymmetric solutions of minimal supergravity in five dimensions'',
\newblock Class.\ Quant.\ Grav.\ {\bf 20} (2003) 4587,
\newblock hep-th/0209114.

\bibitem{gauntlett:0304}
J. Gauntlett and  J. Gutowski,
\newblock ``All supersymmetric solutions of minimal gauged supergravity in
five dimensions'',
\newblock Phys.\ Rev.\ D {\bf 68} (2003) 105009,
\newblock hep-th/0304064.

\bibitem{gutokski:0306}
J. Gutowski, D. Martelli and H. Reall,
\newblock ``All supersymmetric solutions of minimal supergravity in six dimensions'',
\newblock Class.\ Quant.\ Grav.\ {\bf 20} (2003) 5049,
\newblock hep-th/0306236.

\bibitem{caldarelli:0307}
M. Caldarelli and D. Klemm,
\newblock ``All supersymmetric solutions of ${\cal N}=2$ $D=4$ gauged supergravity'',
\newblock JHEP {\bf 0309} (2003) 019,
\newblock hep-th/0307022.

\bibitem{gurrieri:0211}
S. Gurrieri, J. Louis, A. Micu and D. Waldram,
\newblock ``Mirror symmetry in generalized Calabi-Yau compactifications'',
\newblock Nucl.\ Phys.\ B {\bf 654} (2003) 61,
\newblock hep-th/0211102.

\bibitem{cardoso:0211}
G. Cardoso, G. Curio, G. Dall'Agata, D. Lust, P. Manousselis and G. Zoupanos,
\newblock ``Non-Kahler string backgrounds and their five torsion classes'',
\newblock Nucl.\ Phys.\  B {\bf 652} (2003) 5,
\newblock hep-th/0211118.

\bibitem{gauntlett:0212}
J. Gauntlett and S. Pakis,
\newblock ``The geometry of D=11 Killing spinors'',
\newblock JHEP {\bf 0304} (2003) 039,
\newblock hep-th/0212008.

\bibitem{gauntlett:0311}
J. Gauntlett, J.~B.~Gutowski and S. Pakis,
\newblock ``The geometry of D=11 null Killing spinors'',
\newblock JHEP {\bf 0312} (2003) 049,
\newblock hep-th/0311112.

\bibitem{kaste:0301}
P. Kaste, R. Minasian, M. Petrini and A. Tomasiello,
\newblock ``Non-trivial RR two-form field strength and SU(3)-structure'',
\newblock Fortsch.\ Phys.\ {\bf 51} (2003) 764,
\newblock hep-th/0301063.

\bibitem{kaste:0303}
P. Kaste, R. Minasian and A. Tomasiello,
\newblock ``Supersymmetric M-theory compactifications with fluxes on seven-manifolds
and G-structures'',
\newblock JHEP {\bf 0307} (2003) 004,
\newblock hep-th/0303127.

\bibitem{gauntlett:0302}
J.~Gauntlett, D.~Martelli, and D.~Waldram,
\newblock ``{S}uperstrings with {I}ntrinsic {T}orsion'',
\newblock hep-th/0302158.

\bibitem{behrndt:0302}
K.~Behrndt and C.~Jeschek
\newblock ``Fluxes in {M}-theory on 7-manifolds and {G}-structures'',
\newblock hep-th/0302047.

\bibitem{behrndt:0311}
K.~Behrndt and C.~Jeschek,
\newblock ``Fluxes in {M}-theory on 7-manifolds: G-Structures and Superpotential'',
\newblock hep-th/0311119.

\bibitem{Witten:1996mz}E.~Witten,
``Strong Coupling Expansion Of Calabi-Yau Compactification,'' Nucl.\ Phys.\ B
 {\bf 471} (1996) 135, hep-th/9602070.

\bibitem{lukas:9710}
A.~Lukas, B.~A.~Ovrut and D.~Waldram,
``On the four-dimensional effective action of strongly coupled heterotic
 string theory,'' Nucl.\ Phys.\ B {\bf 532} (1998) 43,
hep-th/9710208.

\bibitem{dall:0311}
G.~Dall`Agata and N.~Prezas,
\newblock ``${\cal N}=1$ geometries for M-theory and type IIA strings with fluxes'',
\newblock hep-th/0311146.

\bibitem{martelli:0306}
D. Martelli and J. Sparks,
\newblock ``G-Structures, fluxes and calibrations in M-theory'',
\newblock  Phys.\ Rev.\ D {\bf 68} (2003) 085014,
\newblock hep-th/0306225.

\bibitem{Behrndt:2003ih}K.~Behrndt and M.~Cvetic,
``Supersymmetric intersecting D6-branes and fluxes in massive type IIA string theory,''
Nucl.\ Phys.\ B {\bf 676} (2004) 149 [arXiv:hep-th/0308045].

\bibitem{Behrndt:2004km}K.~Behrndt and M.~Cvetic,
``General N = 1 supersymmetric flux vacua of (massive) type IIA string
theory,'' arXiv:hep-th/0403049.

\bibitem{hackett:0306}
E. Hackett-Jones, D. Page and D. Smith,
\newblock ``G-Structures, fluxes and calibrations in M-theory'',
\newblock JHEP {\bf 0310} (2003) 005,
\newblock hep-th/0306267.

\bibitem{proeyen:9910}
A. Van Proeyen,
\newblock ``Tools for supersymmetry'',
\newblock hep-th/9910030.

\bibitem{acharya:0007}
B. Acharya and B. Spence,
\newblock ``Flux, supersymmetry and M-theory on seven-manifolds'',
\newblock hep-th/0007213.

\bibitem{ali:0111}
T. Ali,
\newblock ``M-theory on seven-manifolds with G-fluxes'',
\newblock hep-th/0111220.

\bibitem{bryant:0004}
R. Bryant,
\newblock ``Pseudo-Riemannian metrics with parallel spinor fields and vanishing
 Ricci tensor'',
\newblock math.DG/0004073.

\bibitem{lu:9805}
H. Lu, C. Pope and J. Rahmfeld,
\newblock ``A construction of Killing spinors on S$^n$'',
\newblock J.\ Math.\ Phys.\ {\bf 40} (1999) 4518,
\newblock hep-th/9805151.

\bibitem{gran:0105}
U. Gran,
\newblock ``Gamma'',
\newblock hep-th/0105086.

\bibitem{horava:9510}
P. Horava and E. Witten,
\newblock ``Heterotic and type I string dynamics from eleven dimensions'',
\newblock Nucl.\ Phys.\ B {\bf 460} (1996) 506,
\newblock hep-th/9510209.

\bibitem{horava:9603}
P. Horava and E. Witten,
\newblock ``Eleven-dimensional supergravity on a manifold with bundary'',
\newblock Nucl.\ Phys.\ B {\bf 475} (1996) 94,
\newblock hep-th/9603142.

\bibitem{salamon} S. Chiossi and S. Salamon, ``The intrinsic torsion of
$\mbox{SU}(3)$ and $\mbox{G}_2$ structures'', Differential Geometry,
 Valencia 2001, World Sci. Publishing, 2002, pp 115-133,
math.DG/0202282.


\end{thebibliography}
\end{document}